\documentclass{article}
\usepackage{sbc-template}
\usepackage{graphicx,url} 
\usepackage{amsmath}
\usepackage{appendix}
\usepackage{amsmath, amssymb}
\usepackage{mathtools}
\usepackage[utf8]{inputenc}  
\usepackage{verbatim}
\usepackage{listings}
\usepackage{xcolor}
\usepackage{bm}
 \usepackage{subfigure}
 \usepackage{natbib}

\usepackage{amsmath}
\numberwithin{equation}{subsection}
\usepackage{amssymb}
\usepackage{graphicx}
\usepackage{booktabs,tabularx}
\usepackage[input-decimal-markers=.]{siunitx}
\usepackage{lipsum}
\usepackage{placeins}
\usepackage{mathrsfs}
\usepackage{appendix}
\usepackage{csquotes}
\usepackage{longtable}
\usepackage{titling}

\begin{document}
\title{A random forest based approach for predicting spreads in the primary catastrophe bond market}
\author{Despoina Makariou, Pauline Barrieu, Yining Chen\\
London School of Economics and Political Science, Statistics Department}
\maketitle
\begin{abstract}
We introduce a random forest approach to enable spreads' prediction in the primary catastrophe bond market.  We investigate whether all information provided to investors in the offering circular prior to a new issuance is equally important in predicting its spread. The whole population of non-life catastrophe bonds issued from December 2009 to May 2018 is used. The random forest shows an impressive predictive power on unseen primary catastrophe bond data explaining $93\%$ of the total variability. For comparison, linear regression, our benchmark model, has inferior predictive performance explaining only $47\%$ of the total variability. All details provided in the offering circular are predictive of spread but in a varying degree. The stability of the results is studied. The usage of random forest can speed up investment decisions in the catastrophe bond industry.\\

\end{abstract}

\section{Introduction}
Catastrophe bonds are \textbf{I}nsurance-\textbf{L}inked \textbf{S}ecurities (ILS), first developed in 1990s, in an effort to provide additional capacity to the reinsurance industry post mega-disasters. The pricing of these instruments is particularly challenging as most of these securities are traded over the counter. Over the last years, there have been several empirical papers trying to address this difficulty by studying the price of catastrophe bonds using real-market data, see \cite{lane2000pricing}, \cite{lane2008catastrophe}, \cite{lei2008explaining}, \cite{bodoff2009analysis}, \cite{gatumel2009}, \cite{dieckmann2010force}, \cite{jaeger2010insurance}, \cite{papachristou2011statistical}, \cite{galeotti2013accuracy}, \cite{braun2016pricing}, and \cite{gotze2018sponsor}. The main orientation of these works was to explain catastrophe bond price via means of identification of variables having a theoretically material and statistically significant link with it. This was vastly achieved through the use of explanatory statistical models. Certainly, the aforementioned works have shed some light on the drivers of catastrophe bond prices. However, there are certain limitations namely selection bias, interactive predictors, fragmented models, non-linearities, and a non-predictive study goal.

Starting from selection bias, the data samples used previously often excluded bonds of certain characteristics, unusual issuances were eliminated as outliers, and observations with missing entries were excluded from data sets, see \cite{bodoff2009analysis}, \cite{gotze2018sponsor}, \cite{galeotti2013accuracy}, \cite{braun2016pricing} and \cite{lane2008catastrophe}. Besides significant loss of information, such study design strategies do not capture that each issuance, in aiming to meet a special risk transfer need, is representative of the whole catastrophe bond population. In \cite{papachristou2011statistical}, concerns about interactive independent variables were expressed but they were not studied. Furthermore, we often see fragmented models accounting only for a given peril territory combination or risk profile, see for instance \cite{bodoff2009analysis}, \cite{papachristou2011statistical} and \cite{lei2008explaining}. From methodological perspective, this is an obstacle in studying the market as a whole especially as new issuances with more rare perils and innovative design arise. From a business perspective, a single model accounting for all transactions is more convenient for model validation ease and flexibility. Another limitation is the extensive use of linear regression without justification of its suitability in a catastrophe bond market setting. This was recognised in some cases, see \cite{lane2008catastrophe} and \cite{papachristou2011statistical}, whilst \cite{major} presented industry based examples of why a simple linear regression model is not appropriate for the catastrophe bond market. Finally, in terms of study goal, past works did not aim at spread prediction, although there is a business need for it, see \cite{major}.

As an attempt to overcome these issues, we suggest a supervised machine learning method called random forest (\citealt{breiman2001random}). Random forest provides highly accurate predictions without over-fitting, see \cite{breiman2001random}, \cite{diaz2006gene}, \cite{oh2003estimating} among others. Most importantly, it is a flexible method in a sense that makes no assumptions about the underlying data generative process tackling the issue of non-linearities and model fragmentation. Moreover, because the building blocks of the method are regression trees, random forest is reasonably robust to outliers and since variables are considered in a sequential manner it captures interactions between variables without the need to specify them (\citealt{breiman1984classification}). Additional advantages of the method are that internal measures of variables' importance can be derived, and selection of the most important variables is feasible. Finally, the need for data pre-processing is minimal, as many steps are integrated in the method itself, ensuring time efficiency from a business perspective. 

In this paper, we apply the random forest method to predict spreads in the full spectrum of primary non-life catastrophe bond market. Our main goal is to generate accurate spreads' predictions of new catastrophe bond observations. The chosen benchmark model is linear regression as it is the predominant model used in the relevant literature. An additional target is to find out whether the details provided to investors in the offering stage of a catastrophe bond issuance are all equally important in predicting its spread. From an empirical viewpoint, we aim at prediction accuracy and variables' importance results to be stable. Given the prediction orientation of our research, we see whether the patterns captured in our data set can provide material for new explanatory-driven studies in the future. Finally, the potential of the introduced machine learning method in facilitating investors' activity in the catastrophe bond market is also of interest. 

We contribute to the literature in the following ways. First, we use a diverse catastrophe bond data set which to the best of our knowledge includes the largest number of data points ever collected before in the primary market setting. Secondly, our study has a purely predictive direction. Thirdly, we use a single algorithmic method to study the market as a whole which differs from the proposition of fragmented models dominating the existing literature.  Finally, we incorporate two variables, i.e. vendor and coverage, that have not been seen in earlier works but they appear in practice as part of the offering circular provided to investors prior a new catastrophe bond issuance.

The rest of the paper is organised as follows. In Section 2, we build on machine learning concepts, in Section 3 we explain our research methodology and in Section 4 we present details about our catastrophe bond data set. Then, the random forest generation based on this catastrophe bond data set is demonstrated in Section 5 whilst the performance of the random forest is evaluated in Section 6. The importance analysis of catastrophe bond spread predictors is found next in Section 7. Furthermore, in Section 8, we provide an example of how the random forest could be used in practice to assist investors' decision making when they examine a new catastrophe bond issuance. Concluding remarks follow in Section 9. 

\section{Machine learning preliminaries}
In this section, we introduce some machine learning concepts that will be useful for the comprehension of methods used later on in our study. The explanations to be given are limited to regression because catastrophe bond spread is a quantitative response variable.

\subsection{Supervised learning}

Machine learning includes a set of approaches dealing with the problem of finding or otherwise learning a function from data (\citealt{james2013introduction}). Supervised learning is a machine learning task where a function, otherwise called a hypothesis, is learned from a data set - often referred to as training set. The latter consists of a number of input-output pairs where for every single input in the training set the correct output is known. An algorithm is going through all data points in the training set identifying patterns and finding how to map an input to an output. Because the desired answer for the output is known, the algorithm modifies this mapping based on how different algorithm generated outputs are compared to the original ones in the training set (\citealt{friedman2001elements}). Ultimately, the aim is that by the time the learning process finishes, this difference will be small enough for the algorithm to be able to map any set of new inputs the algorithm will come across in the future.

\subsection{Ensemble learning}
Sometimes instead of learning one mapping, it is useful to have a collection of mappings which merge their predictions to create an ensemble (\citealt{russell2016artificial}). Individual approximation functions in the ensemble are usually called base learners and predictions combination can happen in various ways with most usual ones being voting or averaging. Such techniques have been investigated quite early on, see for example \cite{breiman1996stacked}, \cite{clemen1989combining}, \cite{perrone1993improving} and \cite{wolpert1992stacked}. The main benefit of ensembles is that if each single hypothesis is characterised by high degree of accuracy and diversity then the ensemble is going to produce more accurate predictions than any of the individual hypotheses on its own, see \cite{zhou2012ensemble}. Here, accuracy means that a hypothesis results at a lower error rate as opposed to one that would be derived from random guessing on new input values while diversity means that each hypothesis in the ensemble makes different errors on new data points (\citealt{dietterich2000ensemble}). Ensembles are usually built by utilising methods to derive various data sets out of the original data set for each base learner. One of the most famous methods to construct an ensemble is briefly discussed below. 

\subsection{Bagging}
Bagging, an acronym for \textbf{b}ootstrap \textbf{agg}regat\textbf{ing} presented by \cite{breiman1996bagging}, is a powerful ensemble learning method. As the name indicates, the ensemble uses the bootstrap, see \cite{efron1992bootstrap}, as resampling technique to take multiple data samples from which multiple base learners will be then generated. At the same time, aggregation, which is simple averaging for regression, is the way to combine the predictions of these individual base learners. There are various merits in using bagging for building ensembles. First, using a bootstrap sample to build each base learner means that a part of the original data (normally two third by default) are not used in its construction. Then, these unseen data points can constitute an unbiased test data to quantify how well each base learner generalises (\citealt{breiman2001random}). Secondly, the method is useful when data is noisy (\citealt{opitz1999popular}).
Thirdly, and probably the most important advantage is that by aggregating base learners which individually suffer from high variance, take decision trees for instance (\citealt{breiman1984classification}), the ensemble as a whole achieves a variance reduction, see \cite{breiman1996bagging}, \cite{bauer1999empirical}, \cite{breiman1996stacked}, \cite{breiman1996heuristics} and \cite{dietterich2000experimental}. A pitfall of the method though is that whilst bagging reduces the ensemble variance, there are diminishing returns in variance reductions. This is because all bootstrap samples are drawn from the same original data set meaning that base learners will inevitably be correlated. This latter point is where the idea of random forest is based on and it will be further discussed in Section 3. 

\section{Research methodology}

Having provided necessary background information about certain machine learning concepts, the purpose of this section is twofold. We start by stating our catastrophe bond spread prediction problem  introducing notations that will be used later in our study. We then continue by presenting our research methodology. 

\subsection{Problem statement with notations}
Broadly, we use an ensemble algorithmic method to perform a supervised learning task for the primary catastrophe bond market. For now, let $\mathbf{x}$  generally denote\footnote{Our convention is that bold lowercase letters reflect random vectors.} the input which reflects characteristics of catastrophe bonds available in the offering circular at the time of issuance. At the same time, let symbol $y$ denote catastrophe bond spreads at the time of issuance. A function $f$ of the form $y=f(\mathbf{x})$ relates catastrophe bond characteristics to their spreads, however $f$ is unknown. Based on past primary catastrophe bond data including information both for $\mathbf{x}=(x_1,x_2,...,x_P)$ where $p=1, 2, ..., P$ and $y$, we first want to find a function that approximates $f$ so that we can predict spreads given new catastrophe bond input.

In particular, experience about past catastrophe bond issuances is captured by collecting $n=1,...,N$ distinct input-output pairs. The input is a vector of predictors, also called features, covariates or independent variables, $\mathbf{x}_n=(x_{1n},x_{2n}, ...,x_{Pn})$ indexed by dimension $p=1, 2, ..., P$ and it is a component of $\mathbb{R}^p$. The output, also called response or dependent variable, is a real-valued scalar denoted by $y_n$ indexed by example number $n=1,...,N$. By assembling these $N$ pairs, we collectively form a catastrophe bond data set $D=\{(\mathbf{x}_n,y_n), n=1, 2, ..., N\}$ based on which the ensemble algorithmic method will search the space $H$ of all feasible functions, in a process called learning, and find a function, denoted by $h_{en}$, that is able to predict the response $y'$ given a new input $\mathbf{x'}$ as  accurately as possible. Because, we use an ensemble method, $h_{en}$ is in reality a collection of functions approximating $f$. We are also interested in assessing the importance of each input of $\mathbf{x}$ in predicting the spread. Finally, all results will be evaluated on the grounds of them being stable, subject to random subsampling of the whole data set.

\subsection{Random forest}

The ensemble method that we use is called random forest. It is developed by \cite{breiman2001random} and here is used to solve prediction problems. As \cite{james2013introduction} mentioned, the underlying logic of random forest is \enquote{divide and conquer}: split the predictor space into multiple samples, then construct a randomised tree hypothesis on each subspace and end with averaging these hypotheses together. Generally, random forest can be seen as a successor of bagging when the base learners are decision trees. This is because random forest addresses the main pitfall of bagging; the issue of diminishing variance reductions discussed earlier in Section 2.3. This is achieved by injecting an additional element of randomness during decision trees construction for them to be less correlated to one another. At the same time, since the base learners are decision trees there are not many assumptions about the form of the target function resulting in low bias. The process of constructing a random forest involves various steps which are summarised in Figure 1 and discussed straight after. 

\begin{figure}[h!]
    \centering
    \includegraphics[scale=0.4]{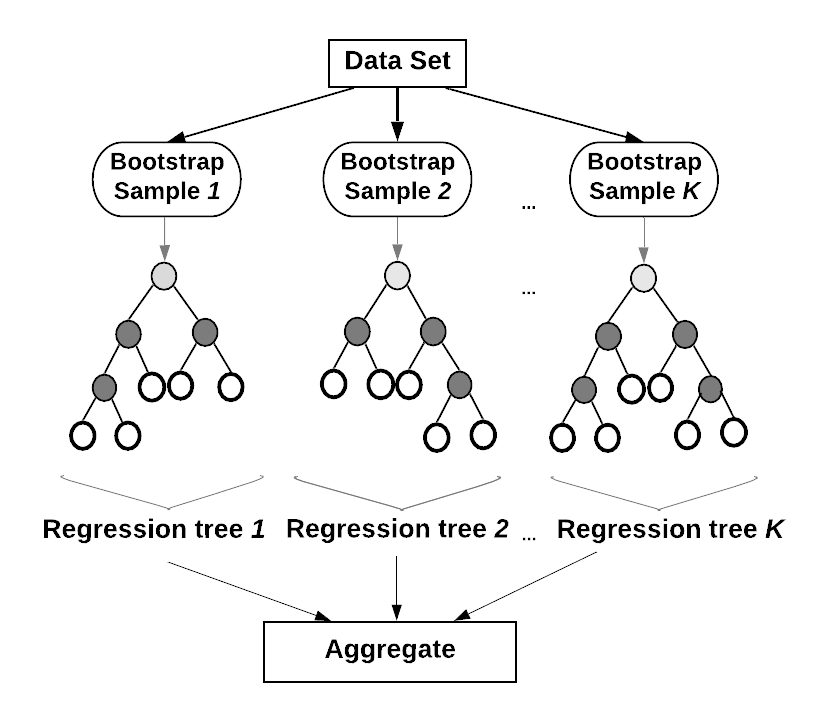}
    \caption{Random forest construction scheme. For each regression tree, light grey circles indicate the root node, dark grey circles intermediate nodes and white colour circles terminal nodes.}
\end{figure}
\FloatBarrier

The first step in the random forest generation process is bootstrap sampling. In particular, from a data set, like $D$, we take $1,..,K$ samples with replacement each of them having the same size as the original data set. The second stage is regression trees development. From each bootstrap sample, $K$ regression trees are grown using recursive partitioning as done in Classification And Regression Trees (CART) (\citealt{breiman1984classification}) but with a smart twist which further randomises the procedure. At each level of the recursive partitioning process, the best predictor to conduct the splitting is considered based on a fresh, each time, random sub-sample of the full set of predictors denoted as $\text{m}_{\text{try}}$. The best split is chosen by examining all possible predictors in this sub-sample and all possible cut-points as of their ability to minimise the residual sum of squares for the resulting tree. A tree stops growing when a minimum number of observations in a given node is reached but generally speaking trees comprising the random forest are fully grown and not pruned. By constructing these $K$ trees we effectively get $K$ estimators of  function $f$ namely $h_1, h_2,...h_K$. The average of these individual estimators $h_{en} = \frac{1}{K} \sum_{k=1}^{K}h_{k}(\mathbf{x}_n)$ is the random forest. 

From the above description, it is evident that there are three parameters whose value need to be fixed prior to random forest development; namely the number of trees grown, node size, and number of variables randomly selected at each split. Each of them respectively control the size of the forest, the individual tree size and an aspect of the within tree randomness. There are certain default values that have been suggested following empirical experiments on various data sets but one can use an optimising tuning strategy with respect to prediction performance to select the most suitable values specifically for the data set under study (\citealt{tunability}). 

After the random forest is built, it can be used to provide predictions of the response variable. To make predictions though, it is necessary to feed the method inputs that have never been seen before during the construction process. As we have briefly mentioned in Section 2.3, due to bootstrap sampling, we can refrain from keeping aside in advance a portion of the original data set for testing purposes. This is why; each tree uses more or less two thirds of the observations, from now on called in-bag observations, whilst the remaining one third of the observations are never used to build a specific tree, from now on called out of bag (OOB) observations. For each tree, the out of bag observations act as a separate test set. To predict the response variable value for the $n^{th}$ observation, one should drop down its corresponding input down every single tree in which this observation was out of bag. This means that by doing so one will end up having in hand on average $K/3$ predictions for any $n=1,...,N$ observation. Then, in order to derive a single response prediction for the $n^{th}$ observation, the average of these predictions is taken. The same procedure is repeated for all other observations. Whether these predictions are good enough or not needs to evaluated based on certain metrics as shown next.

\subsection{Performance evaluation criteria for random forest}
To assess the performance of any machine learning algorithm, one needs to set in advance the criterion upon which judgement will be made. In this paper, we employ two criteria for the performance evaluation of our random forest; prediction accuracy and stability. They are discussed in the two following subsections.

\subsubsection{Prediction accuracy} 
Evaluating our random forest performance based on its prediction accuracy is valuable for the purposes of our study. This is because our goal is to predict catastrophe bond spreads as accurately as possible given new catastrophe bond observations. In general, prediction accuracy is one of the most used performance indicators in machine learning algorithms aiming at prediction. This is no different for random forest algorithm as originally presented in \cite{breiman2001random}. 

In this paper, prediction accuracy is measured by means of the proportion of the total variability explained by the random forest, here denoted as $\text{R}^2_{\text{OOB}}$. Following \cite{gromping2009variable}, the latter metric is defined as $\text{R}^2_{\text{OOB}}= 1- \frac{\text{MSE}_\text{{OOB}}}{\text{TSS}}$ where  $\text{MSE}_\text{{OOB}}$ stands for the total out of bag mean squared error and ${\text{TSS}}$ for the total sum of squares. With respect to $\text{MSE}_\text{{OOB}}$, it shows the variability in the response variable that is not forecasted by the random forest. It is calculated as $\text{MSE}_\text{{OOB}} = \frac{1}{N} \sum_{n=1}^{N}(y_n - \overline{\hat{y}}_{{n}_{\text{OOB}}})^2$ where $\overline{\hat{y}}_{{n}_{\text{OOB}}}$ is the mean prediction for the $n^{th}$ observation where $n=1,...,N$ for all trees for which the $n^{th}$ data point was out of bag. In effect, $\text{MSE}_\text{{OOB}}$ is a sound approximation of the test error for the random forest because every single data point is predicted based solely on the trees that were not constructed using this observation. Actually, when the number of trees $K$ is very large then the $\text{MSE}_\text{{OOB}}$ is roughly equivalent to leave one out cross validation (\citealt{james2013introduction}). Surpassing the need to keep a separate test set is very practical in a catastrophe bond context. This is because one can use all available catastrophe bond data, which is scarce, towards the construction of the random forest to create a stronger prediction method. With regards to $\text{TSS}$, as in linear regression, it reflects the degree at which the response variable, here the catastrophe bond spread, deviates from its mean value. It is defined as $\text{TSS}=\sum_{n=1}^{N}(y_n - \overline{y})^2$ where $y_n$ is the response variable value for the $n^{th}$ observation where $n=1,...,N$ and $\overline{y}$ the mean value of the response variable. In this study, $\text{R}^2_{\text{OOB}}$ is going to be expressed in percentage terms. The higher the $\text{R}^2_{\text{OOB}}$, the better the prediction accuracy of the random forest is. 

\subsubsection{Stability}
The term stability here refers to how repeatable random forest results are when different samples taken from the same data generative process are used for its construction, see \cite{turney1995bias} and \cite{philipp2018measuring} for the rationale behind this approach. The reason for investigating stability is that consistent results are deemed more reliable, see \cite{stodden2015reproducing}, \cite{turney1995bias}, \cite{yu2013stability} and \cite{philipp2018measuring} for a discussion. Moreover, \cite{turney1995bias} had observed that in industry applications, prediction accuracy alone is not enough to gain the trust of practitioners; inconsistent results can create confusion even if the prediction accuracy achieved is high.

Various ways to measure the stability of algorithmic results have been presented in \cite{turney1995bias}, \cite{lange2004stability}, \cite{ntoutsi2008general}, \cite{lim2016estimation} and \cite{philipp2018measuring}. In this study, we are inspired by the works of \cite{turney1995bias} and \cite{philipp2018measuring} with regards to stability and its empirical measurement. In particular, the idea is that by obtaining two sets of data from the same phenomenon sampled from the same underlying distribution the algorithm needs to produce fairly similar results from both data sets for it to be considered stable. One way to achieve this is to randomly partition the whole data set into two separate data sets multiple times. An important decision though is how to take the samples. Here, we propose taking the samples using the split-half technique as described in \cite{philipp2018measuring} meaning that the whole catastrophe data set will be split into two disjoint data sets of roughly equal size. This sampling method ensures that a similarity between the results is not attributed to the same observations being in both samples as this could result in similar results without meaning that the algorithm is actually stable. By choosing a small learning overlap it is possible to examine the degree of a result generalisation for independent draws from the catastrophe bond data generative process. The procedure will be repeated $100$ times.

\subsection{Evaluation of predictors' importance}

The random forest algorithm allows for assessing how important each predictor is with respect to its ability to predict the response, a concept that is briefly called as variables importance. Its assessment is executed empirically (\citealt{gromping2009variable}) and in \cite{chen2012random} one can find a comprehensive review of various methods that can be used to achieve this. Here, the focus lies on two widely used approaches namely permutation importance, and minimal depth importance. 

\subsubsection{Permutation Importance}
The central idea of permutation importance, also known as\enquote{Breiman – Cutler importance} (\citealt{breiman2001random}), is to measure the decrease in the prediction accuracy of the random forest resulting from randomly permuting the values of a predictor. The method provides a ranking for predictors' importance as end result and it is tied to a prediction performance measure. In particular, the permutation importance for ${x}_{p}$ predictor is derived as follows. For each of the $K$ trees; first, record the prediction error $\text{MSE}_{{\text{OOB}}_{\text{k}}}$, secondly, noise up, i.e. permute, the predictor $x_{p}$ in the out of bag sample for the $k^{th}$ tree, thirdly, drop this permuted out of bag sample down the $k^{th}$ tree to get a new $\text{MSE}{_{\text{OOB}_{\text{k}}}^{{{x}_{p}}_{perm}}}$ after the permutation and calculate the difference between these two  prediction errors (before and after the permutation). In the end, average these differences over all trees. The mathematical expression of the above description is $I_{x_{p}}=\sum_{k=1}^{K}[\frac{1}{K}(\text{MSE}{_{\text{OOB}_{\text{k}}}^{{{x}_{p}}_{perm}}}- \text{MSE}_{\text{OOB}_{\text{k}}})]$ where $I_{x_{p}}$ is the importance of variable $x_{p}$, $K$ the number of trees in the forest,  $\text{MSE}{_{\text{OOB}_{\text{k}}}^{{{x}_{p}}_{perm}}}$ the estimation error with predictor $x_{p}$ being permuted for the $k^{th}$ tree, and $\text{MSE}_{{\text{OOB}}_{\text{k}}}$ the forecasting error with none of the predictors being permuted for the $k^{th}$ tree. The larger the $I_{x_{p}}$ the stronger the ability of $x_{p}$ to predict the response. Generally speaking a positive permutation importance is associated with decrease in prediction accuracy after permutation whilst negative permutation importance is interpreted as no decline in accuracy. 

\subsubsection {Importance based on minimal depth}

The other approach for measuring predictors importance is based on measure named minimal depth, presented in \cite{ishwaran2010high} with the latter being motivated by earlier works of \cite{strobl2007bias} and \cite{ishwaran2007variable}. The
minimal depth shows how remote a node split with a specific predictor is with respect to the root node of a tree. Thus, here the position of a predictor in the $k^{th}$ tree determines its importance for this tree.  The latter means that unlike permutation importance, the importance of each predictor is not tied on a prediction performance measure. Also, in addition to ranking variables, the method also performs variable selection - a very useful feature for elimination of less important predictors.

Specifically, \cite{ishwaran2010high} have formulated the concept of minimal depth based on the notion of maximal sub-tree for feature $x_{p}$. The latter is defined as the largest sub-tree whose root node is split using $x_{p}$. In particular, the minimal depth of a predictor $x_{p}$, a non-negative random variable, is the distance between the $k^{th}$ tree root node and the most proximate maximal sub-tree for $x_{p}$, i.e. the first order statistic of the maximal subtree. It takes on values $\{0,...,Q(k)\}$ where $Q(k)$ the depth of the $k^{th}$ tree reflecting how distant is the root from the furthermost leaf node, i.e. the maximal depth (\citealt{ishwaran2011random}). A small minimal depth value for predictor $x_{p}$ means that $x_{p}$ has high predictive power whilst a large minimal depth value the opposite. That said, the root node is assigned with minimal depth $0$ and the successive nodes are sequenced based on how close they are to the root.  The minimal depth for each predictor is averaged over all trees in the forest. \cite{ishwaran2010high} showed that the distribution of the minimal depth can be derived in a closed form and a threshold for picking meaningful variables can be computed, i.e. the mean of the minimal depth distribution. In particular, variables whose forest aggregated minimal depth surpasses the mean minimal depth ceiling are considered irrelevant and thus could be excluded from the model.

\subsubsection{Other evaluation factors}
After calculating the importance of predictors using the methods described above, we consider useful to examine the results based on two additional criteria. First, we want to ensure that the importance rankings and selected variables results are repeatable. Because both permutation and minimal depth importance are linked to the random forest constructed, the stability of predictors' importance results will be evaluated in line with the random forest stability evaluation. Secondly, we will check whether the predictors' importance results reflect investors' knowledge from an empirical perspective.  In a business context, it would be uncomfortable for an investor to see good catastrophe bond predictions but importance rankings of the predictors outside their empirical knowledge, even though from a statistical viewpoint this type of agreement is not necessary. 

\section{Catastrophe bond data}

In this section, we present how the catastrophe bond data used in this study have been collected and processed whilst details are given with respect to the choice of variables and their role in our study. 

\subsection{Collection}
The core of catastrophe bond pricing cross sectional data has been collected from a leading market participant. The websites of ARTEMIS, Lane Financial and Swiss Re Sigma Research have been also extensively used to cross validate data entries that were unclear or non-available in the main data body. To the best of our knowledge, the data collected refer to all non-life catastrophe bonds issued in the primary market from December 2009 to May 2018, a total of $934$ transactions. The information gathered was related to investors return, loss potential of the securitized risk, i.e. expected loss and attachment probability, various design characteristics of the risk transfer, i.e. issuance size, coverage period, coverage type, trigger, region, peril, credit score and risk modelling company. 

\subsection{Preparation}
Consolidating data from various sources was not a straightforward task. In particular, there were pieces of information referring to the same concept but measured in different units across different data providers. For example, some sources expressed expected loss as a percentage of issuance size whilst others in terms of basis points. Since those measured on percentage terms were the majority, the appropriate transformation was made to change the unit from basis points\footnote{The equivalent of $1$ basis point is 0.01 percent.} into percentage terms to maintain consistency within the same data column. With regards to the spread at issuance, it was derived from the coupon by subtracting the element of the money market rate. 

The validation of data across various sources, has been a time consuming task but this was the only way to ensure that there will be no missing values in the study, a pitfall in many previous works. On this note, it needs to be acknowledged that an exception in the above non-missing values claim is few private placement deals where the risk modelling firm was not publicly announced. There, a separate category level was created to capture this specific reason for missingness, i.e. private placement. Including this level is considered important via means that the developed algorithmic method will be able to predict spreads for these circumstances also. Further information on this category level can be found in Appendix A. 

\subsection{Discussion about the choice of variables}
The variables included in the data set can be seen in Table 1, presented along with the definition, type, and their role in this study. In Appendix A, one can find basic statistical information and histograms for all variables along with a discussion to enhance the understanding of catastrophe bond data intricacies. With regards to the role of each variable in our research, the spread was chosen as dependent variable as it is an industry wide accepted lens through which one can see catastrophe bond pricing. The spread is of utmost interest to the investors as it indicates how much they could earn on the top of the risk free rate if they decided to employ their capital in this alternative risk transfer segment. 

\begin{table}[!htbp]
\renewcommand{\arraystretch}{0.4} 
\begin{tabularx}{\textwidth}{@{}lXSS@{}}
\toprule
\multicolumn{1}{l}{Variable}&\multicolumn{1}{l}{Description}&\multicolumn{1}{l}{Type}&\multicolumn{1}{l}{Role}\\
\midrule
spread&The amount of interest earned on the top of the risk free rate.  &{continuous}&{response}\\\\
AP& The probability of incurred losses surpassing the attachment point. & {continuous}& {predictor}\\\\
EL & The annual expected loss within the layer in question divided by the layer size. &{continuous}&{predictor}\\\\
size &Catastrophe bond nominal amount.&{continuous}&{predictor}\\\\
term & Years passed from issuance to maturity date. & {continuous}& {predictor}\\\\
coverage& Contract term indicating whether protection is offered for a string of loss events or a single loss event.  &{categorical}&{predictor}\\\\
diversifier&A peril-region combination.&{categorical}&{predictor}\\\\
rating status & A binary variable showing whether an impartial view of credit quality has been allocated to a catastrophe bond. &{categorical}&{predictor}\\\\
trigger&Mechanism through which a loss payment is activated.&{categorical}&{predictor}\\\\
vendor& Catastrophe risk modelling software firm. &{categorical}&{predictor}\\
\bottomrule
\end{tabularx}
\caption{Catastrophe bond data set dictionary.}
\end{table}
\FloatBarrier

Since the goal of this study lies on the prediction of spread, a major consideration is that the independent variables need to be available at the time of the prediction. This is indeed the case here, as the predictors constitute information included in the placement material offered to investors prior to a new catastrophe bond issuance. To the best of our knowledge, one of the novelties in our study is that we explore the association between coverage type and catastrophe bond spreads. This is in line with current sector discussions as expressed in ILS\footnote{ILS is an abbreviation for Insurance Linked Securities or Insurance Linked Securitisation depending on the context in which it is used.} speciality articles, such as \cite{ILSguide} and \cite{RMS}. There, the need to incorporate the coverage type in catastrophe bond pricing was highlighted following the extensive capital freezes investors experienced after California wildfires in 2018. Briefly touching upon this topic, wildfires, a not well understood peril, has been mostly transferred to investors with a provision that losses are covered on an aggregate basis. By design,  aggregate deals tend to obtain losses easier, even from small events, compared to their per occurrence counterparts, as a string of loss events triggers the bond. The incapacity of the models to account for this to date led to big losses from aggregate deals and pressure for spreads to incorporate this transaction aspect. This signifies the importance of considering this variable. A further addition into the variables kit for studying the spread is the incorporation of information regarding the modelling company employed to calculate the frequency and severity of the securitised catastrophe risks. The software used for this purpose is firm specific thus it is interesting to explore whether by knowing this information part of the spread can be predicted. 

A final note for the variables of this study regards credit ratings. Specifically, the information captured initially was about the actual credit rating being allocated to a given catastrophe bond issuance, if at all. However, here a binary variable was created indicating whether a catastrophe bond issued with a credit rating attached to it or not. This is because we observed that the majority of catastrophe bonds for this period were not rated. Consequently, our focus was shifted from what rating a catastrophe bond has to whether it has any rating at all. The absence of credit rating in new issuances is not solely an observation in the current data set though. In ILS professional circles, the popularity of non-rated catastrophe bonds is justified from a catastrophe bond market evolution perspective; investors feel more comfortable and trust the risk modelling companies for the calculation of loss and the analysis of the risk return profile more and more. As a result, credit ratings are somehow no longer seen as essential as they used to be in the past and this is reflected in the increasing issuance pace of non-rated bonds, see \cite{ARTrate}. In the following sections, we apply the research methodology of Section 3 to the catastrophe bond data set that we have just discussed.

\section{Random forest generation}

In order to build\footnote{The statistical software used is R, version 3.5.1. The statistical packages employed to perform computations are these of; \cite{randomForest} for developing the random forest as well as calculating permutation importance values, \cite{randomForestSCR} for calculating minimal depth importance measures and the minimal depth threshold for variable selection, and \cite{caret} for tuning the main hyperparameter using grid search methodology. It should be mentioned that whenever \cite{randomForestSCR} and \cite{caret} were used, algorithm arguments used agreed to those used in \cite{randomForest} to avoid inconsistencies.} the random forest using our catastrophe bond data set, we first needed to decide the hyperparameters' values that we will use, i.e. number of trees, number of variables randomly selected at each split and node size.  \cite{breiman2001random} has suggested certain default values that seem to work well after multiple empirical experiments however we have incorporated certain tuning strategies for the most important hyperparameters. Our approach in choosing these values are explained below.

\subsection{Number of trees}
The number of trees in the random forest controls its size. Generally, it is good to have a large number of trees as their resulting decisions will be complementing each other more, having a positive impact on random forest prediction accuracy. At the same time, a large  number of trees is a safe option in case the value the optimal value of hyperparameter $\text{m}_{\text{try}}$ is small so that each variable has enough of a chance to be included in the forest prediction process. However, except for the computational cost which is associated with growing large random forests, it was found by \cite{breiman2001random} that there are diminishing returns in the prediction accuracy increase by adding a bigger number of trees. Taking these reflections into account, we have started the random forest development process by growing $2000$ trees and in Figure 2 one can see how the $\text{MSE}_\text{{OOB}}$ converges for various values of random forest size up to this level. From a first sight, it does not take a large number of trees for $\text{MSE}_\text{{OOB}}$ to stabilise. Before even reaching $100$ trees, $\text{MSE}_\text{{OOB}}$ has dropped from above $40000$ to less than $15000$. By the time we reach $500$ trees - a default random forest size value suggested in \cite{breiman2001random}, it seems that the $\text{MSE}_\text{{OOB}}$ has almost been stabilised.

\begin{figure}[!htbp]
  \centering
   \includegraphics[scale=0.5]{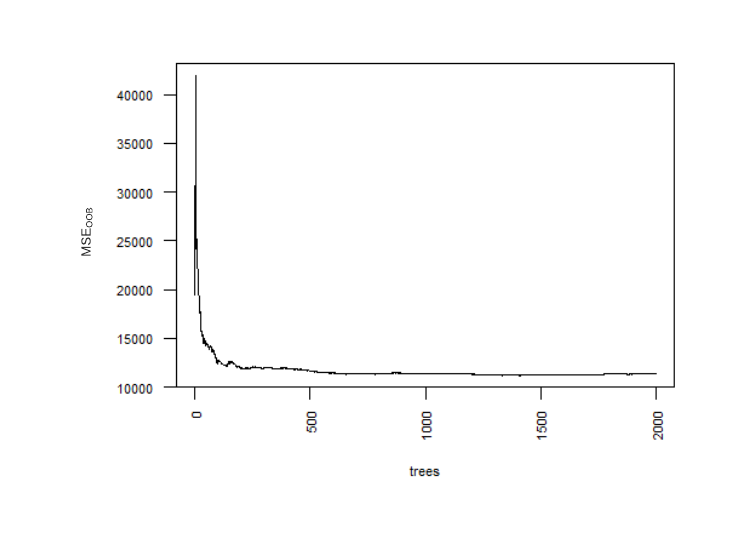}
    \caption{Out of bag mean squared error convergence with respect to random forest size. Mean squared error based on out of bag samples ($\text{MSE}_\text{{OOB}}$) versus number of trees in random forest.}
\end{figure}
\FloatBarrier 

In order to empirically verify that the number of trees we use is satisfactory, we made few extra checks for various numbers of trees below and above the $500$ trees reference point up to $2000$ trees. The results of these computations are shown in Table 2. We see indeed that $\text{MSE}_\text{{OOB}}$ drops further by adding extra trees but it seems that $700$ trees is adequate for our problem. The difference in $\text{MSE}_\text{{OOB}}$ from $700$ trees onward is small and results in the same $\text{R}^2_{\text{OOB}}$. Thus, to avoid extra computational cost, our random forest will be comprised of $700$ trees.  

\begin{table}[!htbp]
\renewcommand{\arraystretch}{1} 
\setlength{\tabcolsep}{40pt}
\centering 
\begin{tabularx}{\textwidth}{l l l} 
\toprule %
Number of trees&$\text{MSE}_\text{{OOB}}$&$\text{R}^2_{\text{OOB}}$\\
\midrule
350&11938.50&92.70\%\\
400& 11928.78&92.71\%\\
500&11676.67&92.86\%\\
600&11491.27&92.97\%\\
\textbf{700}&\textbf{11347.55}&\textbf{93.06\%}\\
800&\textbf{11355.79}&\textbf{93.06\%}\\
2000& \textbf{11351.20}&\textbf{93.06\%}\\
\bottomrule
\end{tabularx}
\caption{Trials for the empirical approximation of random forest size by looking at $\text{MSE}_\text{{OOB}}$ and $\text{R}^2_{\text{OOB}}$ for selected number of trees.}
\end{table}
\FloatBarrier

\subsection{Node size}
The hyperparameter node size controls the size of the tree in the random forest and effectively determines when the recursive partitioning should stop. A large node size results in shallower trees because the splitting process stops earlier. This has the advantage of lower computation times, but it effectively means that that the tree will not learn some patterns resulting in lower prediction accuracy. From the other hand, a small node size translates to a higher computational cost but more thorough learning of patterns and consequently a more accurate base learner.  The recommended value for node size, i.e. the minimum number of data points in the terminal nodes of each tree, given by \cite{breiman2001random}, is $5$ for regression problems. This default value was also suggested and used by many other authors, as \cite{wang2018random}, \cite{gromping2009variable}, and \cite{berk2008statistical} and therefore we also employ it as node size value here. The random forest needs to consist of trees which are fully or almost fully grown, see \cite{breiman2001random}, thus there is not much added value in exploring this aspect further as $5$ meets this requirement and there is a general consensus for its appropriateness.

\subsection{Number of variables selected at each split} 

The number of candidate predictors getting randomly considered at each split, $\text{m}_{\text{try}}$, is the most important hyperparameter. This is because it can affect the performance of the random forest and the predictors' importance measures the most, see \cite{berk2008statistical}. The significance of $\text{m}_{\text{try}}$ lies on the fact that it influences at the same time both the prediction accuracy of each individual tree but also the diversity of the trees in the forest.  To get the most out of the random forest, one wants each tree to have good prediction performance but at the same time trees not to be correlated to one another. However, these two goals are conflicting. An individual tree will be the most accurate when $\text{m}_{\text{try}}$ has a high value but this would result in high correlation for the ensemble. In particular, an extreme case of $\text{m}_{\text{try}}=P$ would force the process to account to simple bagging (\citealt{james2013introduction}). Generally,  a small $\text{m}_{\text{try}}$ is preferable as, for a sufficiently large number of trees, each predictor will have higher chance to get selected and thus contribute in the forest construction. All in all, the trade-off between individual learner accuracy and diversity needs to be managed by finding an optimal value which secures balance for the data set we study.

In \cite{breiman2001random}, the default value of $\text{m}_{\text{try}}=P/3$ is suggested for regression problems. This means that in our problem where $P=9$, the algorithm would consider $3$ predictors at each potential split. We have investigated the relevance of this empirical rule using a tuning strategy called grid search followed by $5$-fold cross validation. The goal was to ensure that the most appropriate $\text{m}_{\text{try}}$ is chosen. The process started by specifying the range of all possible values that $\text{m}_{\text{try}}$ can take, namely the grid. In the current study, this is between $1$ and $9$, i.e. as many as the number of predictors. Then, $9$ different versions of the random forest algorithm were built one for each possible value of $\text{m}_{\text{try}}$. The prediction accuracy of each random forest version, measured by means of $\text{R}^2_{\text{OOB}}$, was evaluated through a 5-fold cross validation.

\begin{figure}[!htbp]
    \centering
    \includegraphics[scale=0.7]{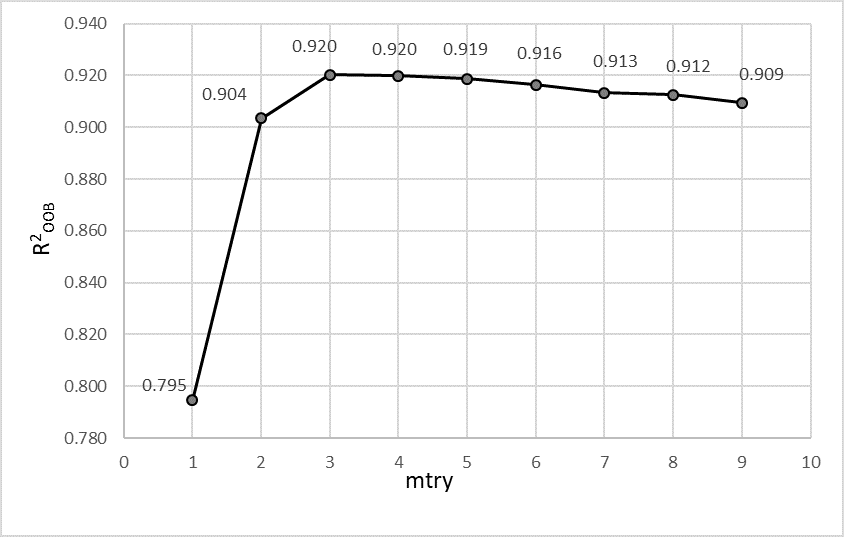}
    \caption{Tuning of main random forest hyperparameter through grid search followed by 5-fold cross validation. Out of bag based $\text{R}^2$ ($\text{R}^2_{\text{OOB}}$) for random forest versus number of candidate predictors getting randomly considered at each split ($\text{m}_{\text{try}}$) during forest generation.}
\end{figure}
\FloatBarrier 

The results, shown in Figure 3, reveal that allowing the random forest algorithm to look randomly at $\text{m}_{\text{try}}=3$ or $\text{m}_{\text{try}}=4$ explanatory variables every time a split is considered at any given node, was almost equally good in terms of prediction performance. However, since variable importance measures were to  be calculated later on, we deemed wise to choose the smaller value of $\text{m}_{\text{try}}=3$, by discipline, as this would lead in less correlated trees giving the opportunity to see the influence of weaker predictors to catastrophe bond spreads prediction. Having decided on the hyperparameter values, the final random forest was generated. The next section investigates how well the random forest performed in our catastrophe bond setting. 

\section{Random forest performance evaluation}
In this section, we evaluate how well our random forest performs with regards to its prediction accuracy and stability. The latter is important as industry users need to feel comfortable that the predictions acquired from random forest are equally good no matter which catastrophe bond data set is used for its construction.

\subsection{Random forest prediction accuracy}

The ability of the random forest to predict catastrophe bond spreads given new input information was investigated. We start by clarifying what we regarded as new input followed by how the catastrophe bond spread predictions were made. Then, the results of the prediction accuracy metrics for the random forest are presented and discussed.

First, as new input for a given tree, we have accounted its out of bag observations. Due to sampling with replacement, only $623$ unique observations, i.e. around two thirds of $N=934$ data points, were used to build each of the $700$ unpruned and almost fully grown (node size = $5$) regression trees. The remaining $311$  observations, i.e. around one third of $N=934$ data points, were never used during the building process for a given tree and as a result they formed a reliable test set for it.  Secondly,  a prediction for the spread at issuance for the $n=1$ observation, $\hat{y}_{1}$, was produced by dropping its corresponding input down every single tree in which the $n=1$ observation was out of bag. This resulted on average to $233$, i.e. around one third of 700, catastrophe bond spread predictions for the $n=1$ observation. Then, a single spread prediction for the $n=1$ observation was made by taking the average  value of these $177$ predictions. After having predicted the catastrophe bond spread value for the observation $n=1$, the same process has been repeated for the rest $n=933$ observations left. Finally, in order to evaluate the prediction accuracy of our random forest, the metrics discussed in Section 3 were calculated. In particular, we have computed the mean squared error based on the out of bag data as $\text{MSE}_\text{{OOB}} = \frac{1}{934} \sum_{n=1}^{934}(y_n - \overline{\hat{y}}_{{n}_{\text{OOB}}})^2$, the total sum of squares as $\text{TSS}=\sum_{n=1}^{934}(y_n - \overline{y}_{n})^2$ and, the variability explained by our random forest as $\text{R}^2_{\text{OOB}}= 1- \frac{\text{MSE}_\text{{OOB}}}{\text{TSS}}$. The results are presented in Table 3 along with additional information about the random forest built.

\begin{table}[!htbp]
\renewcommand{\arraystretch}{1} 
\begin{tabularx}{\textwidth}{@{}lXSS@{}}
\toprule
\multicolumn{1}{l}{Prediction accuracy of final random forest}\\
\midrule
sample size&934\\
number of predictors&9\\
random forest type&regression\\
number of trees&700\\
no. of variables tried at each split ($\text{m}_{\text{try}}$)&3\\
node size&5\\
$\text{MSE}_\text{{OOB}}$&11347.55\\
$\text{R}^2_{\text{OOB}}$&93.06\%\\
\toprule
\end{tabularx}
\caption{Final random forest prediction accuracy results presented in terms of $\text{MSE}_\text{{OOB}}$ and $\text{R}^2_{\text{OOB}}$.}
\end{table} 
\FloatBarrier

It stands out that our random forest explains around $93\%$ of the total variability. At the same time, the predictive performance of our benchmark model, i.e. linear regression, is much lower - it explains only $47\%$ of the total variability \footnote{Details about the prediction performance of linear regression based on the same catastrophe bond data set used for random forest generation are given in Appendix B.}. However, another important aspect in evaluating whether such level of random forest prediction accuracy is high enough, is to consider the nature of the problem under study. On a broader perspective, making predictions in a financial market setting is not an easy task. Inefficiencies, multiple market participants and, the influence of psychology on their behaviour are only few of the factors making the prediction task complex. Consequently, given that we address a financial problem, we could claim that achieving an $\text{R}^2_{\text{OOB}}$ of around $93\%$ here corresponds to a very satisfactory level of prediction accuracy. From a catastrophe bond market perspective though, this result is impressive. Our data set included the whole population of catastrophe bond deals for the $9$ year period under study, with all heterogeneity that characterises it. Effectively, this means that by using the same random forest one can predict the spread of a new catastrophe bond issuance no matter its features and risk profile. To our knowledge this is the first empirical study which achieves to find a single model to account for various types of catastrophe bonds; usually we see a segmentation for US versus non-US perils or earthquake specific models. Using only one pricing solution offers a great flexibility in a business context. An instance of how our random forest can be used in the insurance linked securities industry, is discussed later on in Section 8. In the following subsection, the stability of  prediction accuracy results is assessed.

\subsection{Random Forest Stability}

Here, the random forest results' stability is used as performance criterion and it is measured empirically from a practitioner's point of view as presented in Section 3. Following \cite{turney1995bias} and \cite{philipp2018measuring}, we obtained two sets of data from the same phenomenon and same underlying distribution with as little learning overlap as possible, then constructed two random forests from each one and checked whether prediction accuracy was fairly similar. That said, we took a random $50\%$ of the observations without replacement from the initial catastrophe bond data set, namely Sample A. The rest of the original data set observations, not included in Sample A, formed Sample B. Then, two separate random forests were grown out of Sample A and Sample B to assess the stability of random forest prediction accuracy to changes in the initial data set. We have repeated this process 100 times. Optimal values for the number of variables randomly selected to be considered at each split were sought in both cases. In Table 4, we present a typical realisation of 1 out of 100 iterations with respect to the repeatability of prediction accuracy results.

 Across all 100 iterations, the recorded mean absolute difference of $\text{R}^2_{\text{OOB}}$ between Sample A and Sample B is around $3.6\%$ with the minimum and maximum absolute differences being $0.01\%$ and $10.2\%$ respectively. Given that our problem sits in the intersection of financial and insurance market spheres where many behavioural aspects can affect prices, we consider this difference being small. In essence, it is unlikely that an ILS fund would reject the use of the method solely for such a level of dissimilarity. In fact, the repeatability of prediction results here means that our initial random forest prediction accuracy result, i.e. of an $\text{R}^2_{\text{OOB}}$ of $93\%$ presented in Table 3, is reliable. 

\begin{table}[!htbp]
\renewcommand{\arraystretch}{1} 
\setlength{\tabcolsep}{20pt}
\begin{tabularx}{\textwidth}{l l l}
\toprule
{Random Forest Summary} & {Sample A} & {Sample B}\\
\midrule
sample size&467&467\\
number of predictors&9&9\\
random forest type&regression&regression\\
number of trees&700&700\\
no. of variables tried at each split&3&4\\
node size&5&5\\
$\text{MSE}_\text{{OOB}}$&31132.61&34616.29\\
$R^2_{\text{OOB}}$&80.55\%& 78.87\%\\
\bottomrule
\end{tabularx}
\caption{A typical realisation regarding random forest prediction accuracy stability results.}
\end{table} 
\FloatBarrier

 This finding is beneficial for the usage of the method in the industry. With new catastrophe bonds being issued, the random  forest would need to be validated at some point in time as any other model in an insurance related firm. Surely, in a business context, there is no point in investing time and capital to introduce a new model if the latter provides accurate predictions strictly for one particular data set. Having gone through the examination of prediction accuracy results stability, we proceed with determining the importance of each independent variable in the study.

\section{Predictors importance analysis}
The importance of predictors is assessed using the methodologies of permutation and minimal depth importance presented in Section 3. It should be once again highlighted, that the goal here is to find how powerful each independent variable is in predicting catastrophe bond spreads at issuance. No kind of relationship between spread at issuance and the predictors is to be established - the focus lies solely on their prediction ability. We then compare the stability of predictors importance results for both methods. Finally, for the most stable method, we discuss whether the rankings and variable selection results make empirical sense from investors' viewpoint. 

\subsection{Permutation importance}

The importance of each independent variable in predicting catastrophe bond spreads has been here assessed on the basis of  a percentage increase in
$\text{MSE}_{\text{{OOB}}}$ when a predictor is randomly permuted from the out of bag data whilst others remain untouched. First, the $\text{MSE}_{\text{{OOB}}}$ for each of the $700$ trees comprising the random forest, was recorded. The same process was repeated after randomly shuffling the values of a particular $x_{p}$ across all observations. Then, the change between these two mean squared errors, before and after $x_{p}$ permutation, has been calculated and averaged across the $700$ trees after being normalised by the standard deviations of the differences. This way the importance score for $x_{p}$ has been derived. Finally, based on these scores, an importance ranking has been produced. The ranking of catastrophe bond predictors based on their permutation importance score is shown in Figure 4. Variables higher on the vertical axis are more important in predicting catastrophe spread at issuance.

\begin{figure}[h!]
    \centering
    \includegraphics[scale=0.77]{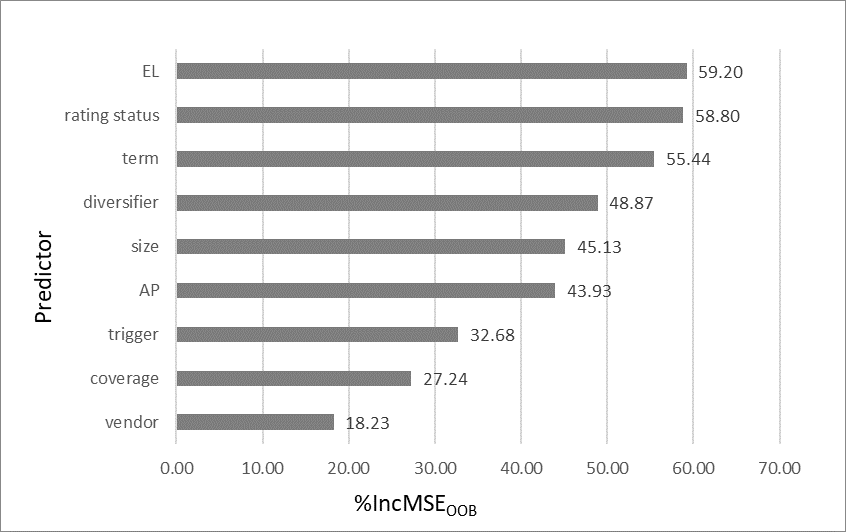}
    \caption{Permutation importance based ranking of predictors. Predictors being permuted versus percentage increase in $\text{MSE}_\text{{OOB}}$ as a result of the permutation.}
\end{figure}
\FloatBarrier 

 One of the first observations is that all scores have positive value. This indicates that each of the independent variables presented here does contribute towards prediction of catastrophe bond spreads. The rating status and expected loss appear as the most important predictors of spread at issuance followed by the variable term. In particular, when the variable rating status is shuffled, the out of bag mean squared error increases by $59.2\%$ whilst the respective percentage for expected loss predictor is marginally lower at $58.8\%$ followed in turn by the predictor term at $55.44\%$. The rest of covariates could be grouped into two different batches with respect to their predictive strength.  Had any of the predictors; diversifier, size and AP been randomly permuted, the prediction performance of the random forest would have been deteriorated between $48.87\%$ and $43.93\%$. Finally, looking at the next batch in the importance ranks, the predictors trigger category, coverage and vendor reduce the random forest forecasting power between $32.68\%$ and $18.23\%$ - a higher percentage range compared to the other two batches of predictors. Next, we present the minimal depth importance results.

\subsection{Minimal depth importance}
The focus is now shifted from using a specific prediction performance measure to assess variables importance to a criterion based on the way that the forest was constructed, i.e. the minimal depth. A tour over the constructed random forest was made to find the maximal subtree\footnote{See Section 3.4.2. for an explanation of what constitutes a maximal subtree.} within each of the $K=700$ trees for a particular $x_{p}$ predictor. From there, the minimal depth for $x_{p}$ within each tree was identified following the rationale explained in Section 3. Then, the forest level minimal depth for $x_{p}$ was derived by averaging the minimal depth for $x_{p}$ within each tree among all $700$ trees. The plot below illustrates the ranking of the covariates with respect to their average minimal depth; higher values of minimal depth correspond to less predictive variables. 

\begin{figure}[h!]
    \centering
    \includegraphics[scale=0.28]{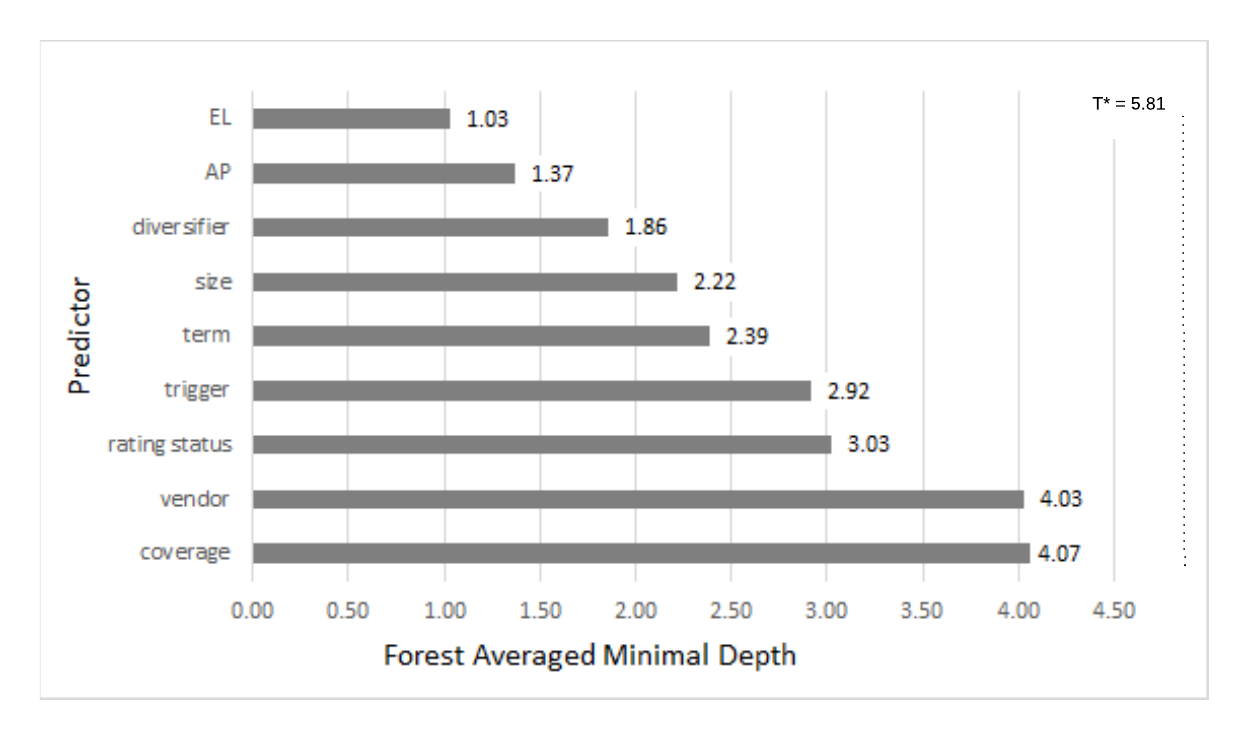}
    \caption{Minimal depth importance based ranking of predictors. Predictors and their forest averaged minimal depth. The dotted line indicates the variable selection threshold, here denoted by T*. }
\end{figure}
\FloatBarrier 

Expected loss and attachment probability with random forest average minimal depths of $1.03$ and $1.37$ respectively have the largest impact in predicting catastrophe bond spreads. In particular, such small values of minimal depth demonstrate that these two variables were mostly used to split either the root node or any of its daughter nodes at least in most of the trees in the forest. Straight after in rankings comes the variable diversifier which on average is chosen to split a node for the very first time at a depth equal to $1.86$. Since this is located close enough to the root node, it is implied that predictor diversifier has also a considerable forecasting power. Using the same rationale, the counterparts; size and term have similar ability to predict catastrophe bond spreads with a marginal distance from their predecessor diversifier, with minimal depth measurements ranging from $2.22$ to $2.39$. It is worth mentioning that among them, the predictor size holds the lead. Next in the importance rankings, one finds the variables trigger and rating status with minimal depth measurements of $3.92$ and $3.03$ respectively. They do have an almost identical minimal depth measurement which can be interpreted as having the same power to forecast catastrophe bond spreads at issuance. Nevertheless, at a random forest level, trigger and rating status are not as powerful because they split nodes which naturally have less data points due to their proximity to the terminal nodes. Finally, vendor followed by coverage type were on average chosen to divide nodes very close to the terminal ones or even the terminal ones. This is revealed by their minimal depth of $4.03$ and $4.07$ respectively, the highest among all predictors, revealing that they have the most limited forecasting ability out of all predictors in the data set.

As mentioned in Section 3.4.2, the minimal depth is not used just for ranking predictors but also for selecting the most important ones. These are those predictors having forest averaged minimal depth lower that the mean of the minimal depth distribution, namely mean minimal depth threshold according to \cite{ishwaran2010high}, here denoted by T*. In the current study, the mean minimal depth threshold equals $5.81$. As seen in Figure 5 though, all predictors had lower forest averaged minimal depth score compared to the one indicated by this threshold. A sensible remark is that all of the predictors in this study carry information which is important in predicting catastrophe bond spreads even though in a varying degree. Although one could use other approaches to select variables, outside the scope of minimal depth, see \cite{chen2012random} for an overview, it is not a necessity here as our data set is not high dimensional, i.e. $P \gg N$. Having presented the predictors' importance results using both permutation and minimal depth based methods, the next discussion refers to the degree of their divergence.

\subsection{Divergence between permutation and minimal depth importance results}

Permutation and minimal depth importance procedures presented for ranking or selecting catastrophe bond spread predictors above are not directly comparable. This is because, as it has been seen, each of them follows a different approach in defining and quantifying the importance in prediction. However, empirically we would expect that there should be some consensus at least for the top and bottom rankings between the two methods.  What we see is that whilst there is a degree of agreement for the least strong predictors, i.e. vendor and coverage, there is considerable divergence at the top and in the middle of the ranks. Given that the low ranks agree, the most worrying difference is for what each method has identified as the most important predictors. This realisation makes us think which of the two variable importance approaches leads to the most trustworthy results for our catastrophe bond spread prediction problem.  Empirically, an answer to this question would be to examine which ranking makes more sense from a practitioner's perspective. However, we believe that first it is preferable to bring our attention back to the concept of results stability but this time for the catastrophe bonds features importance. If one of the two methods is unstable, then we can shift our focus to the one that is most robust and then discuss whether it makes sense from an investor's perspective. That said, the stability checks for the importance results derived by permutation and minimal depth importance methods follow. 

\subsection{Stability checks for predictors importance results}

In this study, a predictors importance method will be considered reliable if its importance ranking for catastrophe bond spread predictors will be fairly robust to data set changes. Shall a change in the catastrophe bond data set from which the random forest is constructed leads to a big change at the top and at the bottom of predictors importance rankings, the variables importance method will be considered unstable and thus probably unreliable. 

Towards this direction, since both permutation importance and minimal depth importance are procedures derived internally after the construction of the random forest, the stability of permutation and minimal depth importance has been mainly examined based on the $100$ random forests pairs grown out of $100$ Sample A and Sample B pairs which have been previously used when the stability of the random forest was investigated in Section 6. In Table 5, we report by variable importance method, the percentage of times where there was an agreement between Sample A and Sample B in the predictor chosen at the top, second, third and bottom positions of the ranking. As bottom positions of the rankings we consider the last two positions jointly. This is because we understand that the further we go down the ranking, variables may be more susceptible to jump from one position to the next or the previous one across different iterations. It is evident that minimal depth importance provides more stable ranking results for the top positions than permutation importance whilst both methods are equally robust for the last two ranking positions.

\begin{table}[!htbp]
\renewcommand{\arraystretch}{1} 
\setlength{\tabcolsep}{20pt}
\begin{tabularx}{\textwidth}{l l l}
\toprule 
Ranking position& Importance method & Agreement percentage
\\ 
\midrule
Top & Permutation & 85\% \\
 & Minimal depth & 96\% \\
\hline
Second from top & Permutation & 27\% \\
& Minimal depth &  84\% \\
\hline
Third from top & Permutation &  27\% \\
& Minimal depth &  70\% \\
\hline
Second from bottom & Permutation & 55\% \\
 & Minimal depth & 38\% \\
\hline
Bottom & Permutation & 58\%\\
& Minimal depth & 43\%\\
\hline
Last two & Permutation & 100\%\\
 & Minimal depth & 100\%\\
\bottomrule
\end{tabularx}
\caption{Ranking stability by predictors importance method.}
\end{table}
\FloatBarrier

 We have also examined which of the two variable importance methods provided the most stable results regarding which variable is chosen at which position of the importance rankings. We considered the number of counts out of $200$ sub-samples taken in 100 iterations (or $400$ samples taken in $100$ iterations when we consider the last two ranking positions jointly), where a given predictor was ranked as top, second from top, third from top or in last two positions in terms of importance by variable importance method. The results are shown in Figure 6 and Figure 7 in terms of percentage frequency. We see that minimal depth is more stable with regards to its predictors choices for the examined ranking positions, as opposed to permutation importance where more variation is visible.

\begin{figure}[!htbp]%
\centering
\subfigure[Permutation Importance]{
\includegraphics[width=7cm]{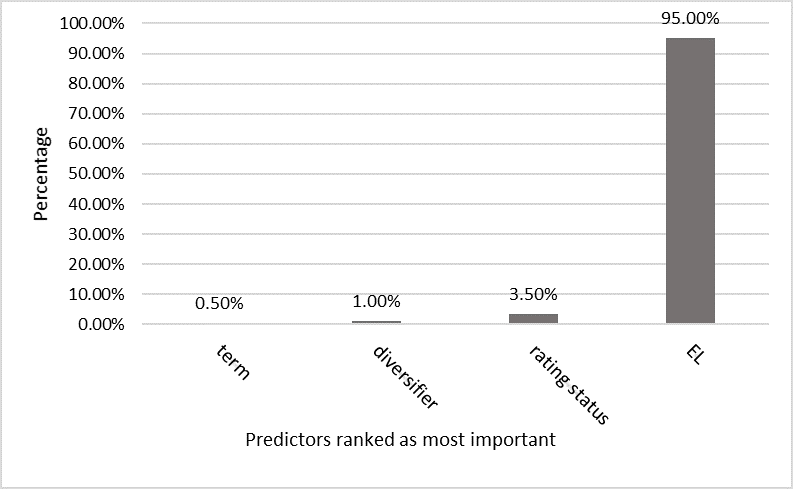}}%
\subfigure[Minimal Depth Importance]{
\includegraphics[width=7cm]{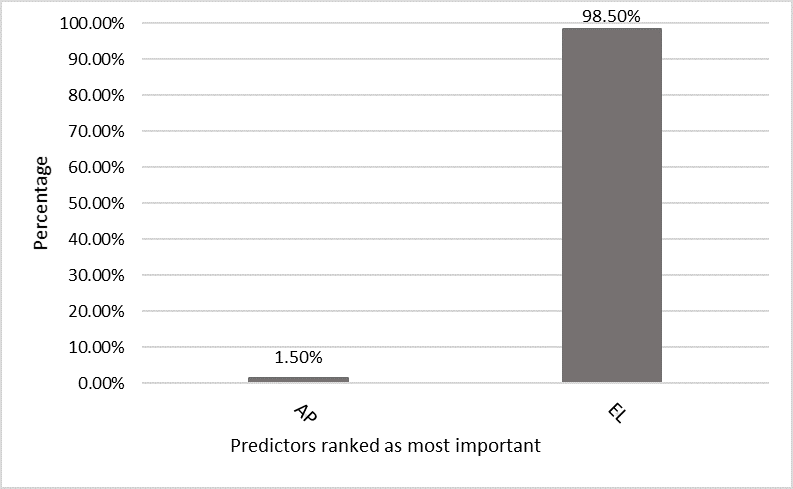}}%

\subfigure[Permutation Importance]{
\includegraphics[width=7cm]{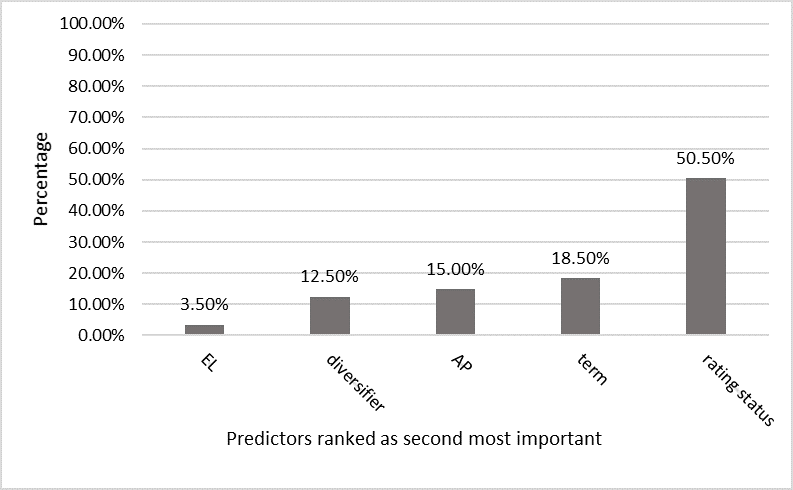}}%
\subfigure[Minimal Depth Importance]{
\includegraphics[width=7cm]{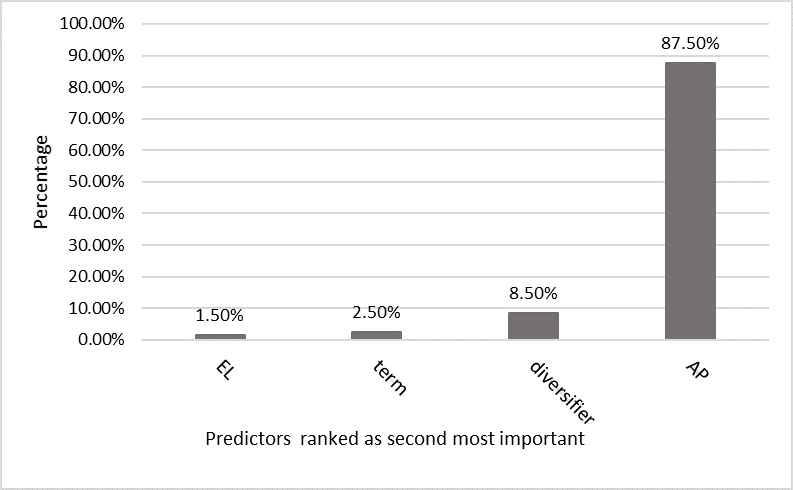}}%

\subfigure[Permutation Importance]{
\includegraphics[width=7cm]{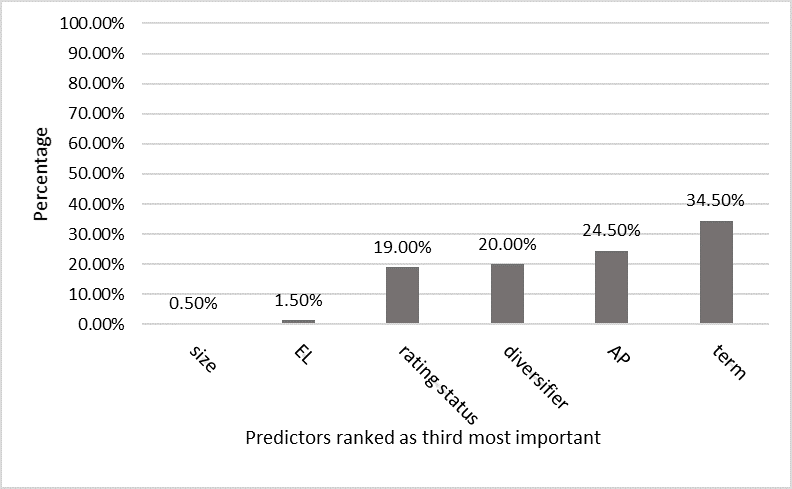}}%
\subfigure[Minimal Depth Importance]{
\includegraphics[width=7cm]{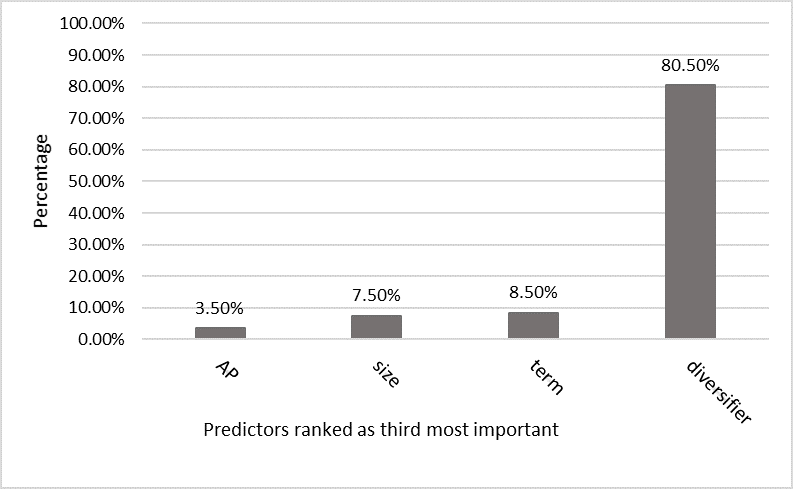}}%
\caption{Bar plots showing the percentage frequency where a given predictor was ranked as top, second from top and third from top in terms of importance by variable importance method.}
\end{figure}

\begin{figure}[!htbp]%
\centering
\subfigure[Permutation Importance]{
\includegraphics[width=7cm]{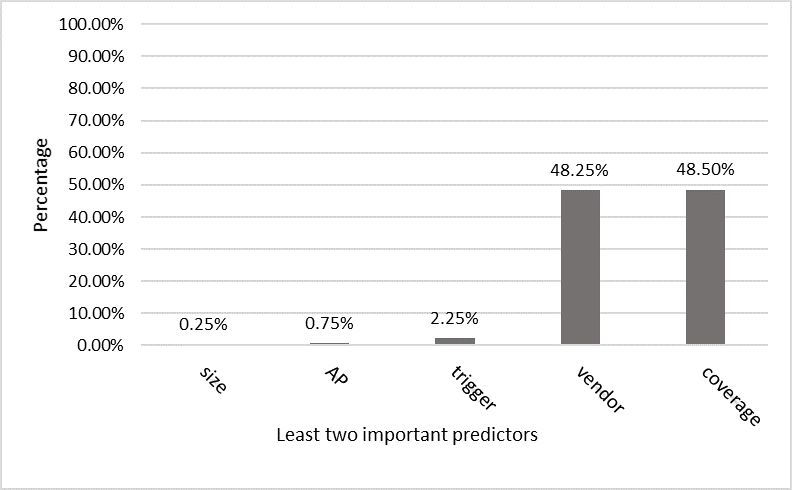}}%
\subfigure[Minimal Depth Importance]{
\includegraphics[width=7.045cm]{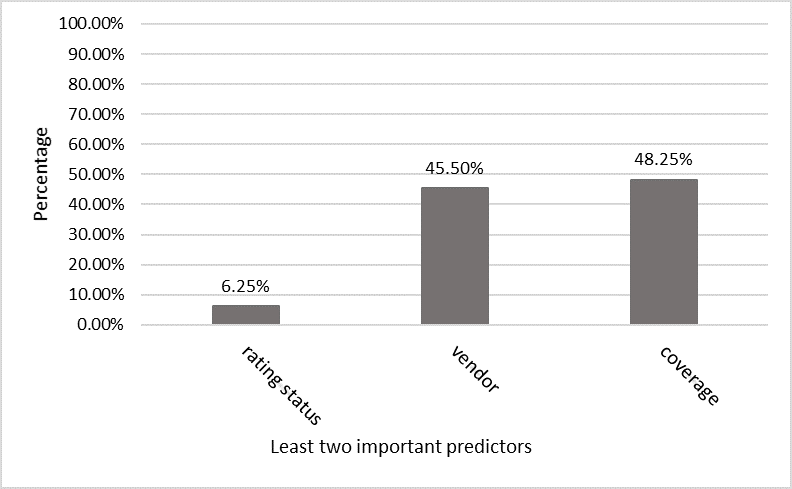}}%
\caption{Bar plots showing the percentage frequency where a given predictor was ranked in the last two positions in terms of importance by variable importance method.}
\end{figure}

As previously highlighted in \cite{chen2012random}, the complex randomisation element of permutation importance procedure makes it difficult to dig any deeper and assess the underlying cause for it being relatively more unstable. However, it should be mentioned that this is not the first work when this measure showed an irregular conduct. As an example coming from the area of bioinformatics, \cite{StabilityRandomForest}) showed that permutation importance rankings were unstable to small perturbations of a gene data set related to the prognosis bladder cancer. All in all, it should be acknowledged that the appropriateness of the feature importance method is mostly data set specific and at least for the catastrophe bond set in hand it seems that permutation importance is not as reliable. Based on the above, any discussion from now on about predictors importance will be based on results of minimal depth importance as presented in Section 7.2.

\subsection{Discussion of predictors' importance results from an industry perspective}

Looking at the minimal depth predictors' importance ranking presented in Figure 5 broadly, we consider three groups of predictive variables, i.e. those of utmost, medium and low prediction strength. We acknowledge that the bounds of where medium and lowest importance variables groups start may be subjective. The distinction here is made looking at the ranking from the perspective of a practitioner. The reason why we want to avoid focusing on individual importance scores is that explaining results in such a detailed way would neither be appropriate nor meaningful for a prediction oriented study. This section is not about interpreting results but seeing whether the results capture somehow investors' perception and knowledge of the market. 
 
Having explained our rationale, the group of top importance predictors comprises from the two fundamental ingredients in any risk quantification process, that is the severity and frequency of losses, i.e. expected loss and attachment probability. This is something that would most probably not surprise insurance professionals, risk managers or even investors if the variable importance results were to be presented to them. Especially with respect to investors, it is well comprehended that the return to be earned by investing into a catastrophe bond deal needs to surpass the amount of payout money should a qualifying event triggers the catastrophe bond payment. Thus, from an empirical viewpoint, investors would expect that by knowing the expected loss and probability of them losing the first dollar, at least a part of the spread value can be predicted.

The second group refers to features which could attract investors' interest in a deal. One reason for this may be the effect that these features have on investors' portfolio returns.  In particular, investors would most probably agree with the predictor diversifier taking the lead in predicting catastrophe bond spreads in the second predictors' group, as this type of information acts as the window shop for them entering the transaction. The rarity of the peril combined with the coverage territory indirectly informs investors about the diversification effect that the particular security can bring into their portfolio; a significant incentive for them to invest in this asset class. We acknowledge that this may not be true for new or rare perils, for which the existing catastrophe models are not yet trusted, however even in this case the peril-territory combination is informative in this sense. Another reason why the predictors of the second group could trigger investment interest is because some of these features are typical in traditional bond types traded in the financial markets with which investors feel more comfortable with. For example, the issuance size, time between issuance and maturity date, credit rating information, and trigger of payment are criteria that any investor would take into account no matter the financial product as they pin point to elements of market's demand for the product, liquidity, credit quality and riskiness. Consequently, one could say that the location of these variables in the ranking supports the way an average investor would think even for a typical non-insurance linked investment. 

Finally, the last group of predictors in the importance ranking comprises from variables having strong technical weight in the securitization process and being insurance sector specific. The first predictor in this group, i.e. vendor, refers to the software company used to calculate the expected loss and various loss probabilities whilst the second one, i.e. coverage type, to a contract term found in insurance contracts. Whilst this may not be immaterial information, there is not direct equivalent of such features in the financial markets. Thus, the average investor not specializing in insurance linked securities would not really dig deep into analyzing vendor model updates and historical catalogs or the wording of the transaction when thinking of what can predict returns. Especially for vendor, it is a matter of fact that there is a global oligopoly in firms offering catastrophe risk modelling solutions in the insurance industry. Although the software developed by each of these companies is based on different assumptions, their scientific grounds are not disputed in the marketplace. This is mostly on the basis that these companies have been founded years before the birth of the first catastrophe bond and also that they have a long track record of being used in the traditional insurance and reinsurance markets. That said there is a contract of trust between them and the market participants as all vendors are perceived to be of equivalent reputational standing. Having said that, it does not mean that investors are sure about the reliability of the expected loss computation. It is just that probably would not believe that one vendor will have a much more valid estimate of loss than another. Similarly, coverage type really matters from an investor's perspective when seen in conjunction with the trigger or the combination of peril and geography. For example, catastrophe bonds with indemnity triggers or not well understood risks when combined with aggregate coverage terms can be risky in trapping investors' capital, as it was seen after 2018 Californian wildfires (\citealt{ILSguide}). However, since the predictors trigger and diversifier were presented earlier in the ranking, the aspect of coverage could be potentially seen as less important in predicting the spread. Taking into account all the above, the minimal depth predictors' importance results seem to reflect investors' current understanding of the market. Next, we discuss at which degree the predictors' importance results agree with past research in empirical catastrophe bond pricing.

\subsection{Predictive versus explanatory importance - links with literature}

It is compelling to see whether the variables that are found in earlier studies to be good at explaining catastrophe bond spreads are any similar to the variables, shown here, to be good at predicting them. As mentioned in \cite{shmueli2010explain}, one should not expect these two to be the same. Variables considered important in explaining the response are tied to a theoretical hypotheses set at the beginning of the study and on the notion of statistical significance. However, these aspects are immaterial in a purely predictive modelling framework as ours. Exploring the level of this divergence is interesting, as it can add value in understanding the full spectrum of catastrophe bond spread drivers for prediction and explanation. It should be mentioned that this is an exercise that shall be made with extra caution as, to the best of our knowledge, every study in the explanatory catastrophe bond pricing literature to date and our predictive study has utilized different data sets and made different assumptions. However, given the fact that some level of agreement has been recorded in the past for certain variables in the explanatory framework, even under these constraints, it is worth having a short discussion.

The starting point is independent variables where harmony with respect to predictive and explanatory importance between this and previous studies has been observed. In particular, in Section 7.2, it was seen that the expected loss is the most major contributor in predicting spreads in the primary catastrophe bond market. This result comes in agreement with almost all explanatory oriented literature where expected loss has been actually included as an independent variable in their study, in particular the works of \cite{lane2000pricing}, \cite{lane2008catastrophe}, \cite{bodoff2009analysis}, \cite{dieckmann2010force}, \cite{braun2016pricing}, \cite{galeotti2013accuracy} and \cite{jaeger2010insurance},  with \cite{lei2008explaining} being the only exception - the (conditional) expected loss was not found to be statistically significant.  At the same time, it was seen that the probability of losses outstripping the attachment point has almost equal forecasting power as the expected loss. This consents with the view of \cite{lane2000pricing} supporting that the catastrophe bond premium is derived through an interplay between frequency and severity of catastrophe bond expected losses. On the top of this, \cite{lei2008explaining} and \cite{jaeger2010insurance} also agreed with the view that the attachment probability is of high significance in explaining catastrophe bond spreads. Furthermore, the peril-territory combination is also of a particular importance as for its ability to forecast spreads here and in the explanatory framework alike results were obtained by \cite{gatumel2009}, \cite{jaeger2010insurance} and \cite{gotze2018sponsor}. Similarly, trigger was found to be predictive in the current research work whilst \cite{dieckmann2010force},  \cite{gotze2018sponsor} and \cite{papachristou2011statistical} have also talked about the explanatory significance of this variable in their models. Finally, the predictor rating status which was found to be predictive in our study (although not of top importance), it was seen as major determinant of spread in \cite{lei2008explaining} and \cite{gotze2018sponsor} even though both works have examined rating from a different perspective to the one we have employed \footnote{In \cite{lei2008explaining}, the variable related to rating captured whether a catastrophe bond has been allocated an investment grade or not rather than whether it has been rated at all or not. Similarly, \cite{gotze2018sponsor} the variable related to rating was not referred to the rating status of the bond but this of the cedent. 
}. 

At a second level, one can see that certain predictors which in past studies were categorised as non-significant spread determinants, here they appear to be relevant for prediction purposes. For example, \cite{papachristou2011statistical} and \cite{braun2016pricing}  had excluded from their analysis the variable term with \cite{dieckmann2010force} being the only one highlighting its importance. At the same time, whilst the predictor size has been minded as less influential or not significant at all by the models of \cite{papachristou2011statistical}, \cite{lei2008explaining} and \cite{braun2016pricing} here it is considered sufficiently important for prediction purposes. This divergence stems from the way weak predictors are treated in a typical linear regression model (like the aforementioned ones) versus random forests. As \cite{berk2008statistical} mentions, in a traditional regression framework a variable having a very small association with the response is most often excluded from the model being regarded as noise. Nevertheless, a big number of small associations when considered not on an individual basis but on an aggregate level can have a substantial impact on fitted values. That is not to say that linear regression is not capable of capturing interactions, however to do so any interactions need to be explicitly specified - a complicated task when the number of predictors in the study starts increasing. On the contrary, random forests, as a tree based method, is naturally able to capture associations between predictors without the need to specify them. Indeed, \cite{papachristou2011statistical} had acknowledged that in the context of his study,  the fact that the term was not considered important to be included in the suggested model could have been due to the challenge of capturing complex effects between covariates. 

Finally, it was seen that the variables vendor and coverage were found to be predictive even though appearing at the bottom of the rankings. However, no link between their predictive versus explanatory performance can be drawn at this point since this is the first time that they were actually studied. A potential reason for lack of prior works taking into account these variables is most probably due to the difficulty of finding information about them as they pin-point to very detailed aspects of a transaction - a view that \cite{braun2012determinants} already expressed regarding the risk modelling company. In the next section, we discuss the application potential of random forest in a real industry setting.

\section{An example of random forest application in the industry}

Given the exceptional predictive performance of random forest for catastrophe bond spreads, we provide an example of how the random forest could be used in a real catastrophe bond market setting. The focus lies on assisting an ILS investor in making faster buying decisions when considering to take on a portion of a newly issued non-life catastrophe bond. Below a generic business problem scenario is presented followed by a random forest implementation which could potentially solve it. 

In particular, just before a new catastrophe bond is issued, potential investors are provided with an offering circular. This document includes information about the deal which is to be launched and an invite for them to attend a road show, post which the issuance pricing will be settled. The information disclosed in this package refers to risk details, various design characteristics of the issuance and a price guidance. Investors want to make sure that the suggested spread compensates them enough for the true element of risk that they would undertake had they entered the transaction.  However, a detailed analysis of this aspect can be time consuming as various departments and sometimes even external risk modelling firms get involved in the process. Whilst this process is undoubtedly important, investors would like to have a first flavour of new deal's potential faster. Then, let's imagine how useful a straightforward prediction tool would be, where investors could plug in details provided in the circular of the new issuance the moment they receive it to get a quick spread prediction for the new transaction they investigate on the spot.

The random forest could do exactly this; after being trained, assessed for the accuracy and stability of its predictions, and used to identify the most important predictors, it could be saved for future use. It is very beneficial for the random forest to be built using information included in the offering material as input. This is because for a new issuance, investors would be provided with the same type of information which in other words would serve as a fresh input. Then, the latter could be simply dropped down the random forest for it to predict the catastrophe bond spread of the new issuance, given the patterns that the random forest has captured between spreads and circular information in the past. This prediction would then be compared with the spread guidance offered and give investors an initial idea on whether the bond is overpriced, under priced or \enquote{fairly} priced based on past catastrophe bond experience. This would direct investors to identify bargains faster and ask more relevant questions about the deal whilst on the road show. Then if the deal is of interest, they could send all information needed to their modelling teams to perform the usual tasks of re-modelling the underlying risk exposure and calculate the marginal impact that this new investment would bring into their portfolio. Overall, random forest is a solution that can speed up the investment decisions and help ILS investment firms not to use their valuable human resources for irrelevant catastrophe bond deals. 

However, one note that needs to be made is that when assessing the discrepancy between predicted spread value provided by the random forest and the price guidance, one might first want to look back at what happened in the past, i.e. the historical discrepancy between predicted and actual values recorded in the prediction phase post the random forest training. This may shed some light on the level at which a mispriced deal according to random forest is due to the portion of variability that the random forest could not explain or merely due to the fact that the new catastrophe bond has characteristics that have never been recorded in the past. The latter problem, could be mitigated if the random forest  would be re-trained at frequent intervals, as part of the model validations taking place each every year in a business context, enriching the training data set with more deals. Given the above, this is just a simplistic example presenting how the random forest could be used in a real catastrophe bond market framework. Although many other parameters could be taken into account for such a new method to be incorporated into internal business processes, here we give an idea of how the prediction power of random forest can liaise with investors' personal judgement to make faster and more informed investment decisions. It should be highlighted that recent developments in the catastrophe risk market also support the use of machine learning techniques. Prime examples are the new cyber risk model of AIR vendor, see \cite{AIRcyber}, and the structuring of the first catastrophe bond relying solely on machine learning for its pricing, see \cite{ledger}.

\section{Concluding remarks}
So far the data-driven catastrophe bond pricing literature was focused on building statistical models with an aim to test causal theory. Due to the heterogeneity of catastrophe bond deals, previous researchers had to focus on a particular segment of catastrophe bond market, mostly a specific peril-territory combination, to explain catastrophe bond spreads. The centre of interest lied on identification of variables which have a theoretically material and statistically significant link to catastrophe bond price, i.e. hypotheses of relationship between price and each independent variable were made. Then a statistical model, mostly linear regression, was applied to observed data to compute the size of this effect and the  statistical significance of each independent variable in relation to the causal hypotheses set at the beginning. For model evaluation, the vast majority of works have used in-sample $R^2$ to assess how well the theoretical model fitted the catastrophe bond data they had in hand. Model selection happened on the basis of keeping statistically significant factors and sometimes those non significant ones having large coefficients to match the function connecting catastrophe bond spread and factors to the true underlying catastrophe bond data generation process. The final results were provided in causal terms; out of those relevant compared to our study, expected loss, diversifier, rating and trigger were found to have a measurable effect on spread with size and term having no effect.  

The approach presented in this research is fundamentally different. A machine learning method called random forest was applied to a rich primary market catastrophe bond data set with a goal to predict the spreads for any type of   catastrophe bond at issuance given the features provided to investors in the offering circular. The word \enquote{any} here signifies that the method handles all bespoke characteristics of catastrophe bond transactions, thus there is no need to silo catastrophe bond observations based on their structure and risk profile and build separate models. Here, we did not focus on the underlying data generation process, instead we learned the association between catastrophe bond spreads and predictors from the data directly using the random forest. The performance of our method was assessed on how accurately it predicts spreads based on unseen catastrophe bond observations. Variable importance measures referred to predictive ability and not the power to explain how the spreads are generated in this universe. There was also interest in securing repeatable prediction accuracy and predictors' importance results thus relevant checks were performed.

It was found that random forest has incredible predictive performance and these results were stable. Moreover, all examined predictors have a say in the prediction of spread even if this is in varying degrees thus they all need to be taken into account. Out of the predictors which were common with those studied before in the literature, predictive and explanatory power coexist for expected loss, diversifier, rating and trigger. The variables size and term were found to have considerable predictive power but in previous studies they were not found to be explanatory. There is potential for random forest to be used in the catastrophe bond industry to fast track investment decisions.

Based on the above findings there are certain aspects that it would be interesting to research in the future. Although by using random forest as presented here, an investor can see whether a new issuance of any type has a competitive price guidance or not, they do not get informed about the suitability of a new deal given their current portfolio composition. Addressing this investors' need is a significant topic for future research. Also, the are few suggestions for the explanatory framework; size and term need to be further investigated as the random forest captured interactions which none of the previous models had attempted before. Also, the explanatory power of coverage type and vendor, which were found to be predictive of spread here, could be also interesting to be studied. All in all, prediction modelling and explanatory modelling may differ but utilising both methods can only increase the understanding of catastrophe bond market segment, increase its transparency and contribute to its development.\\

\Large{\textbf{Appendices}}
\normalsize{\begin{appendix}
\section{Summary statistics for the catastrophe bond data set}
Here, we provide further information about the catastrophe bond data set used in this research paper. Few summary statistics are presented for all variables, both continuous and categorical ones. Starting from the continuous variables, we present histograms in Figure 10 and measures of central tendency and spread of the observations in our data set in Table 6. In Figure 10, we see that all continuous variables have a right skewed distribution with only exception variable term. In particular, term distribution has two peaks reflecting that most catastrophe bond issuances have a 3 to 5 year time horizon. Looking at Table 5, we notice that the range between minimum and maximum values for all continuous variables as well as the interquartile range are rather broad indicating that data points are well spread out. Such a data structure is anticipated in a catastrophe bond market setting. In essence, each issuance is a bespoke product developed to meet a very specific risk transfer need and consequently the population of catastrophe bond deals is heterogeneous. A worth mentioning point is that the minimum value of spread in our data set is zero. In contrast to the dominant view that low spread associates to low insurance risk assumed by investors, this is not the case here. In particular, the zero spread observations happen to be related to risky catastrophe bond tranches which were sold to investors at a discount, as part of Residential Re 2017 and Laetere Re 2016 1 Series issuances, and carried a zero coupon. This example signifies once again the diversity of catastrophe bond transactions; each one is inherently different from another.

\begin{figure}[!htbp]%
\centering
\subfigure[Histogram for response spread]{
\includegraphics[width=5cm]{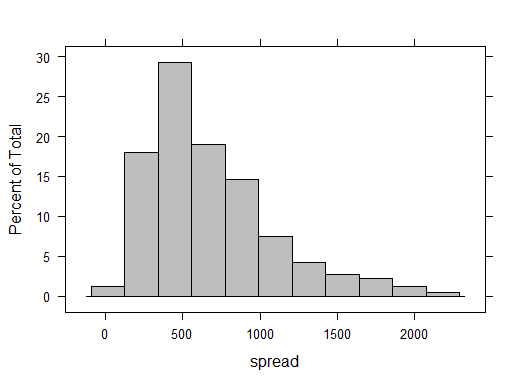}}%
\subfigure[Histogram for predictor EL]{
\includegraphics[width=5cm]{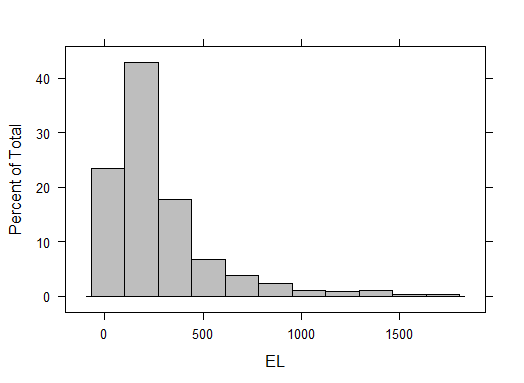}}%
\subfigure[Histogram for predictor AP]{
\includegraphics[width=5cm]{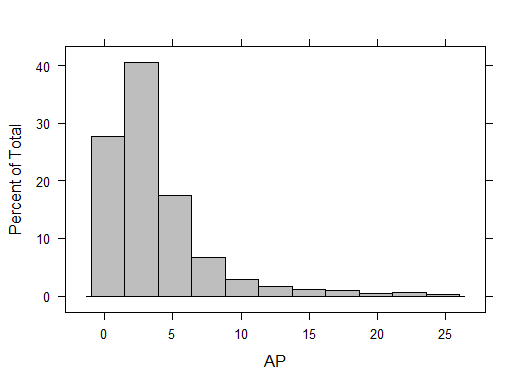}}%

\subfigure[Histogram for predictor size]{
\includegraphics[width=5cm]{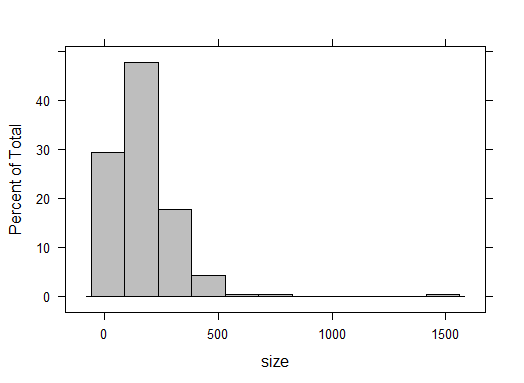}}%
\subfigure[Histogram for predictor term]{
\includegraphics[width=5cm]{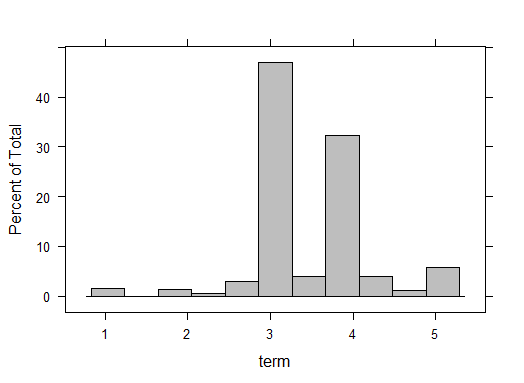}}%
\caption{Histograms for continuous variables. Percent of total observations versus a numerical variable.}
\end{figure}

\begin{table}[!htbp]
\renewcommand{\arraystretch}{1} 
\centering 
\begin{tabularx}{\textwidth}{l l l l l l l } 
\toprule %
Continuous Variable& Min. & 1st Qu. & Median & Mean & 3rd Qu. & Max.
\\ 
\midrule
spread (as \% of size) & 0.00 & 3.75 & 5.75 & 6.75 & 8.60 & 22.00\\
EL (as \% of size) & 0.01 &  1.11 &1.88 &2.75 &3.34 &17.35 \\
AP (\%)& 0.02 &  1.36 &2.51 &3.72 &4.68 &25.04 \\
size (in million US dollars) & 3.00& 75.00 & 130.00 & 164.70 & 200.00 & 1500.00\\
term (in years) & 1.00 & 3.02 & 3.18 & 3.49 & 4.02 & 5.12\\
\bottomrule
\end{tabularx}
\caption{Continuous variables summary statistics. The unit in which each continuous variable is measured is provided in brackets.} 
\end{table}
\FloatBarrier

Moving forward to categorical variables in Table 7, we present for each of them the number of level and number of observations under each level, with the latter quantity also being expressed as a percentage of the total number of observations. All variables levels are those used by the industry unless otherwise stated. Some comments regarding each categorical variable follow.  With regards to coverage type, we find that the majority of catastrophe bonds during the studying period were issued to provide compensation in situations where a single large-scale loss event  would activate the trigger, i.e. per occurrence coverage, as opposed to this happening due to a collection of insured loss events i.e. aggregate coverage. In very few instances in the data set, such as tranches A and B of Riverfront Re Ltd Series 2017-1 for example, per occurrence and annual aggregate coverage co-existed. 

With respect to diversifier, we shall start by providing some explanations in terms of abbreviations. APAC stands for perils specific to Asia Pacific region, NA for perils relevant to North America, SA for prominent perils in South America, Europe for perils in the aforementioned region, and Multi peril for various peril-territory combinations. We see that more than half of the catastrophe bond deals in the data set had a mixture of perils in various geographical territories as an underlying. This is justified to an extend because the diversification effect in these instances is much higher making this type of transactions more attractive to investors.  Bonds covering wind in North America follow in terms of popularity even if the assumption of losses in the area is more likely due to the effect of hurricane seasons. Nevertheless, the high frequency of events had allowed risk modelling companies to understand the risk better, and build more trustworthy models with investors feeling more secure to buy exposures in this region. Looking into the rating status of the bonds issued, it is evident that more than half of catastrophe bonds in the data set did not receive a rating by any independent credit quality agency.

With regards to triggers, indemnity ones were the most popular among the bonds included in the study followed by industry indices. This clearly shows a preference from cedents' perspective to get compensated for the exact level of losses that they anticipate to experience or at least to be compensated in line to industry losses. Deals which are triggered when pre-determined event parameters are satisfied or surpassed accounted only for $5.6\%$ of the total market in the period under study. Examples of parametric index deals in the current data set is Atlas VI Capital Ltd. Series 2010-1 and Bosphorus Ltd. Series 2015-1 whilst IBRD CAR 118-119  is an example of pure parametric trigger deal issued by the International Bank for Reconstruction and Development for Mexico's natural disaster fund named FONDEN. The least used triggers were those combining different trigger types such as Fortius Re II Ltd. Series 2017-1 and those based on the modelled losses of the cedent's exposure portfolio  calculated based on event parameters gathered from specified agencies, such as Akibare II Ltd. single tranche. 

{\small 
\begin{table}[t]
\renewcommand{\arraystretch}{1} 
\centering 
\begin{tabularx}{\textwidth}{l l l l} 
\toprule %
 Categorical Variable & Levels & No. of Observations & Percentage
\\ 
\midrule
{coverage} & {aggregate} &303 &32.4\\
&  {occurrence} & 627 &67.1\\
\bottomrule
{diversifier} & {APAC} & 73 & 7.8\\
&  {Europe} & 66&7.1\\
&  {Multi Peril} & 528 & 56.5 \\
&  {NA Quake} & 80 & 8.6\\
&  {NA Wind} & 184 &19.7\\
&  {SA Quake} & 3 & 0.3\\
\bottomrule
{rating status} 
&  {rated} & 435&46.6\\
&  {not rated} & 499&53.4 \\
\bottomrule
{trigger}
&  {Indemnity} & 511 & 54.7\\
&  {Pure parametric} & 29 &3.1\\
&  {Industry loss index} & 325 &34.8\\
&  {Parametric index} & 23 &2.5\\
&  {Model} & 22 &2.4\\
&  {Multiple} & 24 &2.6\\
\bottomrule
{vendor} & {AIR} & 741 & 79.3\\
&  {AON} & 4 &0.4\\
&  {EQECAT} & 42 &4.5\\
&  {RMS} & 141 &15.1\\
& {pp} & 6 & 0.6\\
\bottomrule
\end{tabularx}\\
\caption{Summary statistics for categorical variables. Levels of each categorical variable are presented by number of observations and percent of total observations. Abbreviations are explained in the text.} 
\end{table}
\FloatBarrier
}

With respect to the risk modelling company used to calculate the expected loss of investors' exposure to underlying peril, we see that AIR Worldwide is the most widely used followed by RMS. Together, they account for the $94.4\%$ of all non-life securitisations in the data sample followed by EQECAT, AON and pp accounting for the rest $5.6\%$. It is worth to note that pp abbreviation is not a risk modelling firm but it stands for \textbf{p}rivate \textbf{p}lacement. Examples are the single tranches of Merna Re Ltd. Series 2016-1, 2017-1, 2018-1 which were privately purchased by specialized ILS funds. Finally, the internal model of AON was used for very few deals where the aforementioned company had acted as the structuring and placement agent, such as in the case of Windmill I Re series 2013-1.

\section{Random forest versus linear regression prediction accuracy}
We believe it is important to compare the accuracy of catastrophe bond spread predictions derived using the random forest as opposed to those derived using a benchmark model. Given that most of the previous literature used   literature segment, the most usual is mostly the traditional linear regression model. To this direction, we built a linear regression model using our catastrophe bond data set focusing on model's prediction performance. For consistency reasons, the bootstrap was one of the resampling methods tried out to estimate the prediction accuracy of the linear regression model. This was done as it was described earlier in the case of random forest; we used $700$ bootstrap samples to refit the model and for each observation, we only considered predictions from bootstrap samples not including that observation. Then, the prediction accuracy as measured by means of out of bag $\text{R}^2$ in the linear regression case was $47\%$ percent as opposed to $93\%$ in the case of random forest. Similar prediction accuracy results for the linear regression case were derived using $10$-fold cross validation with $\text{R}^2=51\%$ and leave one out cross validation with $\text{R}^2=47\%$. Based on the above, we see that random forest significantly outperforms linear regression for catastrophe bond spread prediction purposes.

\end{appendix}}

\normalsize{\bibliographystyle{agsm}}
\bibliography{ref}
\end{document}